\documentclass[12pt,preprint]{aastex}
\usepackage{emulateapj5}
\begin {document}

\title{The Magnetic Field Structure of the LMC 2 Supershell: NGC 2100}

\author{John P. Wisniewski\altaffilmark{1,2,3}, Karen S. Bjorkman\altaffilmark{3,4}, 
Antonio M. Magalh\~aes\altaffilmark{3,5}, Antonio Pereyra\altaffilmark{5}}

\altaffiltext{1}{NASA GSFC Code 667, Greenbelt, MD 20771 USA, jwisnie@milkyway.gsfc.nasa.gov}

\altaffiltext{2}{NPP Fellow}

\altaffiltext{3}{Visiting Astronomer, Cerro Tololo Inter-American Observatory}

\altaffiltext{4}{Ritter Observatory, Department of Physics and Astronomy MS 113, University of Toledo, Toledo, OH 43606, karen.bjorkman@utoledo.edu}

\altaffiltext{5}{IAG, Universidade de S\~ao Paulo, Caixa Postal 3386, S\~ao Paulo, SP 01060-970, Brazil, mario@astro.iag.usp.br, antonio@astro.iag.usp.br}

\begin{abstract}

We present U,B,V,R,I imaging polarimetry of NGC 2100 and its surrounding
environment, which comprise a part of the LMC 2 supershell.  The morphology 
of the observed position angle distribution provides a tracer
of the projected magnetic field in this environment.  Our polarization
maps detail regions exhibiting similarly aligned polarization position angles,
as well as more complex position angle patterns.  We observe regions of
coherent fields on spatial scales of 42 x 24 pc to 104 x 83 pc, and infer
projected field strengths of $\sim$14-30$\mu$G.  We propose that the superposition of global 
outflows from the LMC 2 environment, as well as outflows created within NGC 2100, 
produce the unique field geometry in the region.  

\end{abstract}


\section{Introduction}

Magnetic fields are known to align dust grains, hence influence astrophysical processes, in 
a wide variety of environments (see e.g. \citealt{goo96,laz03}).  The alignment of grains 
in the diffuse interstellar medium of 
the Milky Way \citep{ma70a} and the Magellanic Clouds \citep{ma70b} has long been known 
from linear polarization studies, which measure the dichroic absorption of starlight.  Similarly, 
polarimetric observations indicate grains may also be aligned in environments such 
as dark clouds \citep{laz97,hil99}, and the 
circumstellar environments of young \citep{ait93,tam99} and old \citep{ait95} stars.
As reviewed within \citet{laz03}, a number of alignment mechanisms have been proposed 
over the past sixty years, and it is likely that the local conditions of each 
astrophysical environment will dictate the relative importance of each of these 
proposed alignment mechanisms.  Characterizing the typical polarimetric 
behavior in each of these environments offers one avenue to constrain the relative 
importance of these mechanisms in each environment.   

Massive stars in OB associations provide a rich source of
outflows from stellar winds and supernovae explosions; the interaction of
such outflows with the surrounding interstellar medium (ISM) is known to
create large supershells.  Observations suggest that some of these shells 
may be magnetized, with typical fields strengths of at least several to tens of $\mu$G 
\citep{val93,val94,per07}, and it is expected that these fields can influence the 
dynamical evolution of shells, including constraining their expansion 
\citep{min93,tom98}.  With a diameter of $\sim$ 900 pc,
LMC 2 was initially identified as a supershell based upon its
morphology \citep{mea80}.  The
kinematics of the LMC 2 supershell has been a subject of some debate: 
\citet{cau82} suggested that LMC 2 was expanding as a cohesive structure; 
however, followup studies by \citet{mea87}, \citet{poi99} and \citet{amb04} 
argue against a 
global expansion, asserting the structure is a conglomeration of localized
expanding structures.  Observed extended X-ray emission in the region led
\citet{wan91} to suggest that LMC 2 was formed as a result of a
superbubble breaking out from the plane of the Large Magellanic Cloud (LMC).  

The richest of the 29 stellar associations in the direction of LMC 2 
identified by \citet{hod67} projected to lie interior
to LMC 2 is NGC 2100. Given its age, $\sim$15 Myr \citep{cas96}, 
\citet{poi99} has postulated that all of NGC 2100's O stars have exploded as
supernovae, implying that it has played an important role in shaping the
observed structure of LMC 2.  \citet{poi99} and
\citet{amb04} observed complex H$\alpha$ velocity components which vary across
the extent of NGC 2100, suggesting the presence of a turbulent interstellar
environment.  The interstellar polarization near NGC 2100 is also known to differ 
from general behavior of the LMC (0.32 - 0.57\% at position angles of 28-45$^{\circ}$, 
\citealt{wi05a,wi05b}); \citet{ma70b} observed a strong focusing of magnetic lines 
around 30 Doradus (30 Dor), which is located $\sim$16$\farcm$ west of the LMC 2 supershell.  

In this paper, we present detailed polarization maps of NGC 2100 and its surrounding 
environment, which reveal evidence of a complex magnetic field morphology.  These 
data provide diagnostics of the grain alignment mechanisms which likely dominate in 
such dynamic astrophysical environments.  In Section 2, we
describe our polarimetric observations.  Polarization maps and estimates
of projected magnetic field strengths are presented in Section 3.  In
Section 4, we present a discussion of these results and the clues they
may offer towards understanding the formation and evolution of the 
LMC 2 supershell.

\section{Observations}

Imaging polarimetry of NGC 2100 and its surrounding field were
obtained at the CTIO 1.5 m telescope, as summarized in Table 1.  We used
the F/7.5 secondary configuration, yielding a 15$\farcm$0 field of view and
a 0$\farcs$44 pixel$^{-1}$ scale.  Data were recorded with the telescope's
standard Cassegrain focus CCD (CFCCD), a 2048 x 2048 CCD which was
read out in dual amplifier mode.  The standard telescope configuration was
modified by the addition of a rotatable half-wave plate, followed by a
dual calcite block (Savart plate) placed in the first filter wheel.  This dual calcite 
analyzer, whose optical axes were crossed to minimize astigmatism and color
effects, simultaneously produced two orthogonally polarized images of 
all objects, allowing for the near complete cancellation of all background
sky polarization, as well as atmospheric transparency effects \citep{mag96}.  
A small imperfection in the calcite
block was masked out, shrinking the effective field of view of the instrument
by 2$\farcm$0-3$\farcm$0 arcminutes in the southeast corner of the CCD and 
by $\sim$1$\farcm$5 in the 
northeast and northwest corners of the chip.  Standard Johnson U, B, V, R, 
and I filters
were housed in the second filter wheel.  Images were taken at 8 wave-plate
positions, each separated by 22.5$^{\circ}$, allowing us to derive full
linear polarization measurements for NGC 2100.  Additional information
regarding this instrument can be found in \citet{mag96}, \citet{mel01}, 
and \citet{per02}.

Basic image processing was done in IRAF\footnote{IRAF
is distributed by the National Optical Astronomy Observatories, which are
operated by the Association of Universities for Research in Astronomy, Inc.,
under contract with the National Science Foundation.} using standard 
techniques.  After deriving aperture photometry for our images, the least 
squares solution of the 8 wave-plate positions, calculated with the
PCCDPACK polarimetric reduction package \citep{per00}, yielded linear
polarization measurements.  The residuals at each wave-plate position, $\psi_{i}$, with 
respect to the expected $\cos 4\psi_{i}$ curve constitute the uncertainties in 
our data; these are consistent with the theoretically expected photon noise
errors \citep{mag84}.

Instrumental polarization effects were determined from observations
of polarized and unpolarized standard stars, obtained nightly during our
ten day observing run in 2001.  These data are self-consistent and
agree with observations obtained in a subsequent eleven day observing
run in 2002 \citep{wis03} using the same instrument, illustrating 
its excellent stability.  The instrumental polarization was measured to be within 
0.03\% (I filter) to 0.07\% (B filter), thus no correction was applied to
our data.  Note that due to a lack of known faint unpolarized standard stars,  
we used the M star HIP 21556 as an unpolarized standard star for the 
B, V, R, and I filters.  Given its nearby location (d=11 pc) and spectral type,
one would not expect such an object to exhibit significant 
polarization \citep{tin82}.
Indeed, observations of this target during our 2001 and 2003 observing runs
showed it to be unpolarized. 

\section{Results}

The polarization data we discuss represent the superposition of two components
of distinctly different origin, namely interstellar and intrinsic polarization.
Intrinsic polarization can arise from a variety of scattering mechanisms
within the circumstellar environment of a host star, while interstellar 
polarization results from dichroic absorption of starlight by aligned
interstellar dust grains located along the line of sight.   
Most of our targets should be normal main-sequence stars which are not characterized 
by the presence of an extended (or asymmetrical) circumstellar envelope, hence 
they will exhibit no intrinsic polarization at the precision level
of our measurements.  The statistical analysis of the total polarization observed 
for each of our targets should thus provide an accurate diagnostic of the interstellar 
polarization along the line of sight \citep{mcl79,per02,wis03,wi05b}. 
Furthermore, given the LMC's large distance of 50 kpc \citep{fea91}, any
spatial variability we detect in this interstellar polarization component must  
be the result of a change in the magnetic field or interstellar dust grain
properties within the LMC, rather than a projection of Galactic interstellar
medium properties.

\subsection{Polarization Maps}

In Figures 1-5 we present the polarization in the U, B, V, R, and I filters
for NGC 2100 and its surrounding field.  Polarization vectors are overplotted
on Digitized Sky Survey2 (DSS2) red (V, R, and I filters) and blue (U and B
filters) images which span 0.5 square degrees, allowing one to place the
polarization of NGC 2100 in the context of the LMC 2 supershell.  A more
detailed image of LMC 2's nebulosity, including identification of the major OB 
associations in the area, can be seen in Figure 1c of \citet{poi99}.  
To exclude likely spurious detections, we have only plotted objects
with the following properties: $ 0.1\% <$ polarization $< 3.0\%$ 
and $p/\sigma_{p} > 3.0$.  

Numerous trends in the morphology of the polarization vectors in
each of the filters can immediately be seen.  The magnitude of a typical
polarization vector is $\sim 1.5\%$, which is significantly higher than the
average polarization observed throughout the LMC \citep{wi05a,wi05b}.  
While polarization position angles (PA) tend to be coherent
on small spatial scales, large-scale variability across the field of view
of the data set is immediately apparent.  The patterns traced by this 
large-scale variability are consistent across every filter, indicating the
phenomena represent real features.  

To further explore these large-scale position angle trends, we divided
our field of view into 5 smaller spatial scales which each seemed to possess
one unique, average position angle.  We assign the arbitrary labels A-E
to these fields, and show these fields in the R filter in Figures 6-10.  Note that we have 
only plotted all objects with $ 0.1\% <$ polarization $< 3.0\%$
and $p/\sigma_{p} > 3.0$ in these figures.
The mean position angle, FHWM of gaussian
fits to the samples, standard deviation (in radians) of objects within the gaussian
fit for each area, and spatial extent of each area, assuming a distance of
50 kpc, are tabulated in Table 2.  The polarization position angle histograms used to 
derive these parameters are also given in Figures 6-10.

The position angle rotation to the north, west, and south of NGC 2100, 
apparent by casual inspection of Figures 1-5, 
is indeed real as we measure the mean PA to vary from 
76$^{\circ}$ to 122$^{\circ}$ to 94$^{\circ}$ in Figures 7-9 respectively.
Figure 10 depicts spatial area E, immediately west of NGC 2100, showing
PA alignment at 167$^{\circ}$.  The dramatic curvature in the field pattern
traced out in areas B-E in the area west of NGC 2100 
qualitatively matches a similar ``bubble-like'' pattern seen in the 
H$\alpha$ image of the LMC 2 region of \citet{poi99}, e.g. their Figure 1a.

Systematic alignment in the eastern portion
of our field of view is less dramatic, as seen in Figure 6 which depicts
area A.  We find position angles in area A tend to be $< 90^{\circ}$, 
and find suggestive evidence that the PA distribution in this spatial region
may be fit by two gaussians at 29$^{\circ}$ (FWHM = 25$^{\circ}$) and
64$^{\circ}$ (FWHM = 22$^{\circ}$) respectively.  The constituents of these
two possible gaussian
distributions occupy no unique spatial regions: it is possible that the
observed distributions originate at slightly different distances within the
LMC 2 neighborhood, projecting themselves onto common spatial regions.
Alternatively, it is possible that large-scale LMC-2 flows to the west of NGC 2100 are 
interacting with velocity fields from within NGC 2100 to produce the observed apparent 
superposition effects. 

The projected spatial extent of regions with definitive, coherent position 
angle alignment varies from
42 x 24 pc, associated with region E, to 104 x 83 pc, associated with region
B.  We note the suggestive presence of smaller alignment trends within
and outside of some of these designated areas.  Due to small number 
statistics, it is unclear whether such features are real, hence illustrating
common alignment on finer scales, or whether they are the
 result of small intrinsic
polarization components adding small scatter to the data set.  Deeper
polarimetric mapping of the region would clarify the presence of small-scale
alignment by providing a larger statistical database.  If our polarization
maps are indeed tracing some of the larger structures in the H$\alpha$ maps
of \citet{poi99}, then we would expect deeper polarization maps to trace
many of the finer nebulosity seen in such images.

\subsection{Estimating B Fields}

A formalism for estimating magnetic field strengths from polarimetric
observations was developed by \citet{cha53}(C-F technique), and has been since modified
to account for various inadequacies 
(see e.g. \citet{goo96,zwe96,hei01,cru04}).  As summarized by \citet{hei01,hen01} and \citet{per07}, while the C-F method is a commonly used technique to estimate magnetic 
field strengths, its use is dubious in cases of, amongst other 
factors, large polarization position angle dispersions and large turbulent velocity 
dispersions.  We have used the description
provided by equation 7 of \citet{hei01}, \begin{equation} 
B = \frac{1}{2}\sqrt{4 \pi \rho \left( \begin{array}{c} \frac{\sigma(v_{los})^{2}}
{\sigma(tan \delta)^{2}} \end{array} \right)}
\end{equation}, where $\rho$ is the mean density, $\sigma(v_{los})$ is the
dispersion in the line-of-sight velocity, and $\sigma(tan \delta)$ is the dispersion in polarization position angles, i.e. the difference, within the distribution, between the 
position angle of a given object and the average position angle.   This 
description was used as
it eliminates the small angle approximation present
in the original formalism; furthermore, it includes a factor of 1/2 to account for 
the field overestimation provided
by the classical Chandrasekhar-Fermi method \citep{cru04}.

We were able to measure position angle dispersions for regions B-E of our
dataset using PCCDPACK and tabulate the standard deviation of these values, 
$\sigma(tan\delta)$ in Table 2.  \citet{poi99} reported a
HI number density of 3-4 cm$^{-3}$: we assumed a number density of 4
cm$^{-3}$ in our calculations.  We estimated the line of sight velocity for our 
regions from the HI velocities reported by \citet{mea87} for their region 43, 
corresponding to the approximate location of NGC 2100, $\sigma_{vlos}$ = 
52 km s$^{-1}$.  This dispersion is consistent with the FWHM of H$\alpha$ 
velocities reported by \citet{poi99} across their E-I spectroscopic cut.  
The resulting
magnetic field strengths for our fields range from 14-30 $\mu$G, as tabulated
in Table 2.  We stress that these field values should only be considered
crude estimates: detailed measurements of the gas velocity dispersions and
densities corresponding to the specific spatial regions in which we observed
polarization position angle dispersions are needed to further refine these
field estimates.  Nevertheless, our derived range of field strengths are consistent with the
strength of random field fluctuations in the LMC reported by \citet{gae05}, especially 
those located nearby supernova remnants and wind bubbles, which were quoted to 
be $\sim$8 $\mu$G by these authors.

\section{Discussion}

We now explore some of the implications of the polarization maps presented in
Section 3.1.  Our polarization maps of NGC 2100 and its nearby environment 
indicate the
presence of a sizable magnitude of interstellar polarization, $\sim1.5\%$
which experiences systematic position angle changes.  We attribute this
position angle variability to changes in the orientation of the projected
magnetic field.  It is equally likely that the third dimension of this
field also varies.  The only effect such a variation would have on our data set
would be additional dispersion in the distribution of polarization levels.  
While a wide distribution is observed, 
other factors such as the presence of small intrinsic polarization components 
in objects, small variability in the distance of objects within our 
field of view,
and changes in the polarizing properties of the grains across the field of 
view, i.e. changes in grain size, shape, or composition, also likely serve
to broaden the observed distribution.  Additional observational tools, 
such as that provided by atomic alignment \citep{yan06}, could be used 
to provide an independent measure of the localized three-dimensional 
magnetic field.  

\subsection{Origin of Position Angle Variations}

We consider the origin of the position angle variability detailed
in Section 3.1.  Given the dynamic nature of the region,
it seems plausible to expect the complex field patterns, which align grains to
produce the observed polarization, to be driven by the various outflows 
present.  Such a scenario is supported by the possible tracing of large 
H$\alpha$ features by our data, as noted in Section 3.1.  The morphological
details of Figures 1-5 suggest that the field patterns are not solely
guided by the stellar outflows and supernovae remnants of NGC 2100.
Studies of the interstellar polarization
surrounding other young LMC clusters and OB associations with similar hot
star contents do not illustrate these types of complex field 
patterns \citep{wi05a,wi05b}.  Within the LMC 2 environment, the western side of
NGC 2100 shows complex field patterns while the eastern side of the 
cluster only displays moderate evidence of cohesive field alignment.  No
asymmetry in the distribution of massive stars or their remnants in this
cluster has been observed, thus we don't expect winds or outflows from 
NGC 2100's massive star population to be responsible for producing these
field patterns.  Rather, we speculate that other large-scale flows might play a major role 
in twisting field lines in the observed patterns.

\citet{poi99} suggest that rather than being a cohesively expanding shell, 
the geometry of LMC 2 is that of two HI sheets enclosing a region of hotter
gas.  They suggest both
cavity material and the surface of the HI sheets are being swept eastward across
the complex by the outflows of material located on the western edge of the
region.  We speculate that such general, large-scale flows, carrying
with them local magnetic field lines, might move past
the northern and southern boundaries of NGC 2100 and be impeded by the cluster
itself.  Such a scenario could
account for the coherent position angle patterns located to the north and south
of NGC 2100, as well as the dramatic turnabout in the field immediately west
of the cluster.  As the neighborhood due east of NGC 2100 would be partially
shielded from such flows, one would expect less coherent field patterns
in this environment, as is observed to the east of NGC 2100.
The 30 Dor complex, a rich site of powerful stellar outflows, is
located to the west of NGC 2100.  From polarization measurements, 
\citet{ma70b} noted strong magnetic 
focusing in the 30 Dor region.  Thus we suggest 30 Dor should be considered
as a possible source of outflows which influence LMC 2 and shape the 
projected magnetic field patterns we observe.

A number of grain alignment mechanisms have been postulated and, as summarized in 
the review paper of \citet{laz03}, it is likely that different mechanisms may dominate in different astrophysical environments, depending upon the local conditions present.  
Some of the proposed mechanisms include 
the Davis-Greenstein process, in which paramagnetic dissipation by rotating 
grains leads to alignment 
\citep{dav51}, the Gold process, in which grains are mechanically aligned via 
collisional interactions with a supersonic gas flow \citep{gol52,laz94,la97b}, 
and radiative torques, in which alignment is achieved via the spin-up of irregularly shaped grains which scatter left- and right-hand polarized light in a different way \citep{dol76,dra96,dra97}.  The LMC-2 supershell itself, if it assumed to be a cohesively 
expanding body (e.g. \citealt{cau82}), is only characterized by an expansion velocity of $\sim$30 km s$^{-1}$ \citep{cau82}, which is well below the supersonic gas velocity 
required for the Gold mechanical alignment mechanism.  However,  
NGC 2100's proximity to both the 30 Dor region and the winds of massive stars within 
NGC 2100 suggest that local grains might interact with a more dynamic gas flow than 
that which characterizes the much larger LMC-2 region.  As such, we suggest that 
mechanical alignment might indeed play a partial role in constructing the observed 
morphology of aligned grains in the NGC 2100 region; clearly detailed modeling of 
the system would be advantageous to quantitatively constrain the various grain alignment 
mechanisms which could be operating in this dynamic environment.

\subsection{Location of Polarizing Region}

While we have interpreted the bulk of the magnetic field variability implied
by our observations to be tied to the dynamics of the inner layer of LMC 2, 
we now consider the possible influence of the HI sheets which encompass 
this layer in
the proposed 3-dimensional picture of \citet{poi99}.  Based upon the
derived total line of sight reddening for NGC 2100, $E_{B-V}$ = 0.24 \citep{kel00}, the standard
relationship between polarization and extinction presented in \citet{ser75},
$P_{max} <$ 9 E$_{B-V}$, predicts an interstellar polarization of $< 2.2\%$, in
agreement with the average magnitude observed in our data set, $\sim1.5\%$.
\citet{poi99} report the thin (80-100 pc) HI sheets have 
column densities of $\sim$1 x 10$^{21}$ cm$^{-2}$.  This column density implies
a reddening value similar to that of \citet{kel00}, hence a similar predicted
maximum magnitude of interstellar polarization, based upon the relation
$N_{HI} / E_{B-V}$ = 5 x 10$^{21}$ atoms cm$^{-2}$ mag$^{-1}$ \citep{sav72}.
Thus it appears that enough dichroic absorption by interstellar dust grains
could occur within the thin HI sheet positioned in front of NGC 2100 to
produce the level of observed polarization.

\subsection{Summary}

We have presented polarization maps for a subsection of the LMC 2 supershell,
namely NGC 2100 and its surround field.  These maps show regions of aligned
position angles on scales of 42 x 24 pc to 104 x 83 pc, attributable
to absorption by interstellar dust grains aligned by projected magnetic
fields.  We estimate these projected fields to have strengths of 8-17
$\mu$G, and stress that more accurate field estimates may be achieved by
incorporating measurements which better reflect the interstellar medium
properties corresponding to our survey area.  A plausible explanation for the
observed complex field patterns is that outflows present within LMC 2, modified by 
velocity fields from NGC 2100, combine to produce the observed field patterns. 
The observed asymmetrical field morphology suggests the
stellar sources in NGC 2100 are not the primary source of outflows shaping
the observed fields.  Rather, we speculate that NGC 2100 may serve to 
disrupt the path of large-scale flows moving eastward across LMC 2.  
We suggest that the 30 Dor region, observed to be a source
of both massive outflows and strong magnetic fields, may be the source 
powering the observed field patterns in LMC 2.

Finally, we considered a proposed 3-dimensional picture of LMC 2 in which two  
HI shells confine a region of hotter gas.  We find the magnitude of observed
polarization could be produced by aligned dust grains within
one of these HI shells, noting that some mechanism must then impart the
complex field geometry produced within the inner gas layer to this thin outer
shell.

\acknowledgments

We thank the anonymous referee whose comments helped to improve this 
paper.  This research was supported by NASA NPP and GSRP fellowships to JPW  
(NNH06CC03B, NGT5-50469), a NASA LTSA grant (NAG5-8054) and a 
Research Corporation Cottrell Scholar award to KSB, and a FAPESP grant 
(02/12880-0) to AP.  AMM also acknowledges support from the Brazillian 
agencies FAPESP and CNPq.  Polarimetry at the University of S\~ao Paulo (USP) 
is supported by FAPESP.  This  
research has made use of NASA's Skyview virtual observatory, NASA ADS, and
the SIMBAD database.

\clearpage
%\newpage
\begin{table}
\caption{Journal of NGC 2100 Observations}
\begin{tabular}{lcc}

Filter & Obs. Date & Exposure Time \\

\tableline

U & 24 Nov. 2001 & 1200 sec. \\
B &  23 Nov. 2001 & 240 sec. \\
V & 23 Nov. 2001 & 180 sec. \\
R & 24 Nov. 2001 & 180 sec. \\
I & 23 Nov. 2001 & 180 sec. \\

\tablecomments{Note that the listed exposure times 
correspond to the total integration at each of 8 wave-plate positions.}

\end{tabular}
\end{table}

\clearpage
%\newpage
\begin{table}
\caption{Summary of Magnetic Field Properties by Region}
\begin{tabular}{lcccccc}

Region & \# Stars & Spatial Extent (pc) & Mean PA (deg) & FWHM (deg) & $\sigma(tan\delta)$ (rad) & B ($\mu$G) \\

\tableline

A & 105 & 73 x 177 & 0-90 & \nodata & \nodata & \nodata \\
B & 110 & 104 x 83 & 76 & 24 & 0.24 & 25 \\
C & 88 & 63 x 75  & 122 & 37 & 0.26 & 23 \\
D & 56 & 104 x 63 & 94 & 22 & 0.20 & 30 \\
E & 56 & 42 x 24  & 167 & 46 & 0.44 & 14 \\

\tablecomments{Summary of the polarization position angle variability
across our field of view.}

\end{tabular}
\end{table}

\clearpage

\begin{figure}
\includegraphics[scale=0.5]{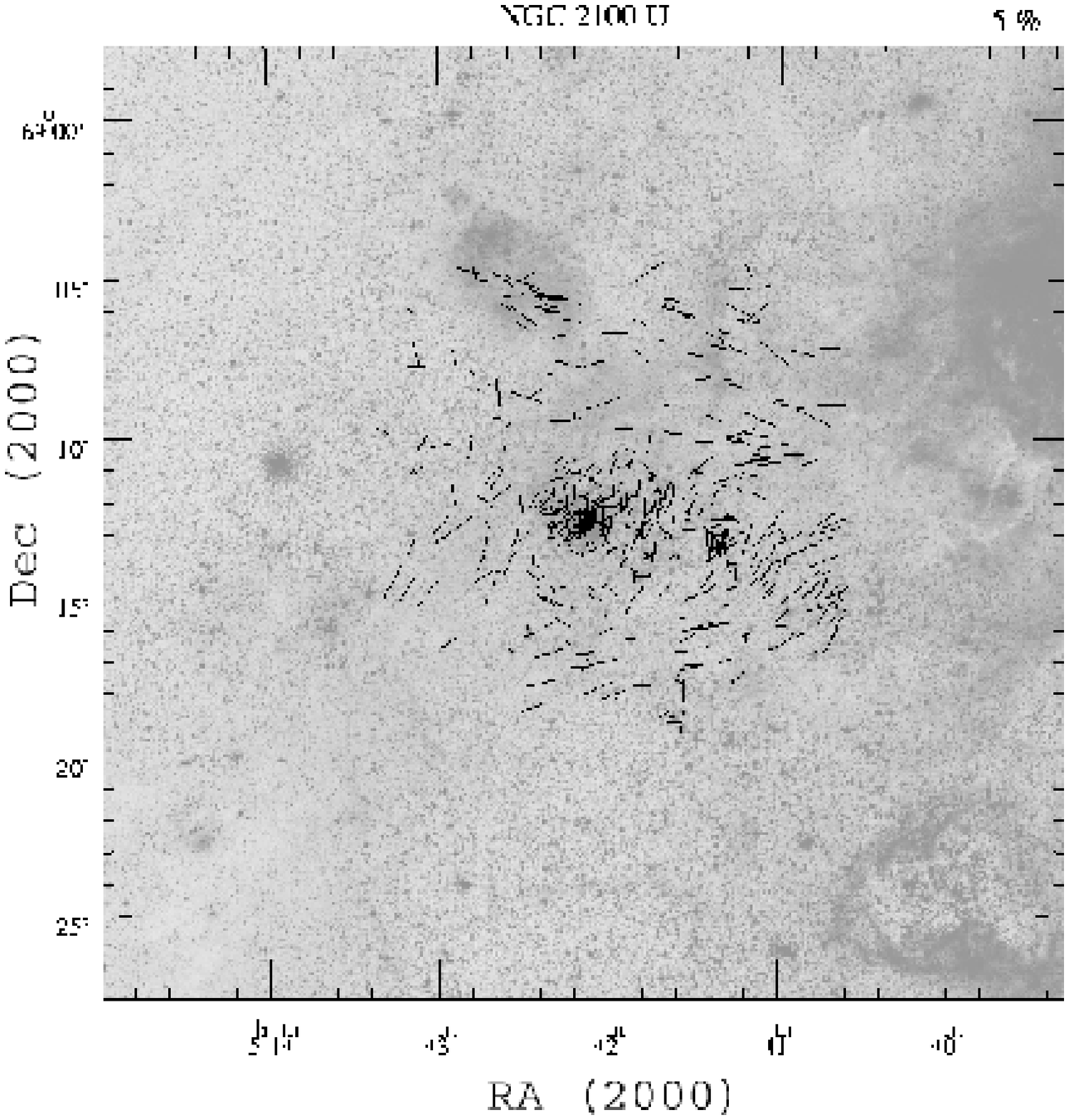}
\caption{U filter polarization of 462 targets which have 
p$/\sigma_{p} > 3$ in the vicinity of NGC 2100.  Polarization vectors
are overplotted on a 0.5 deg$^{2}$ DSS2 blue image.}
\end{figure}

\begin{figure}
\includegraphics[scale=0.5]{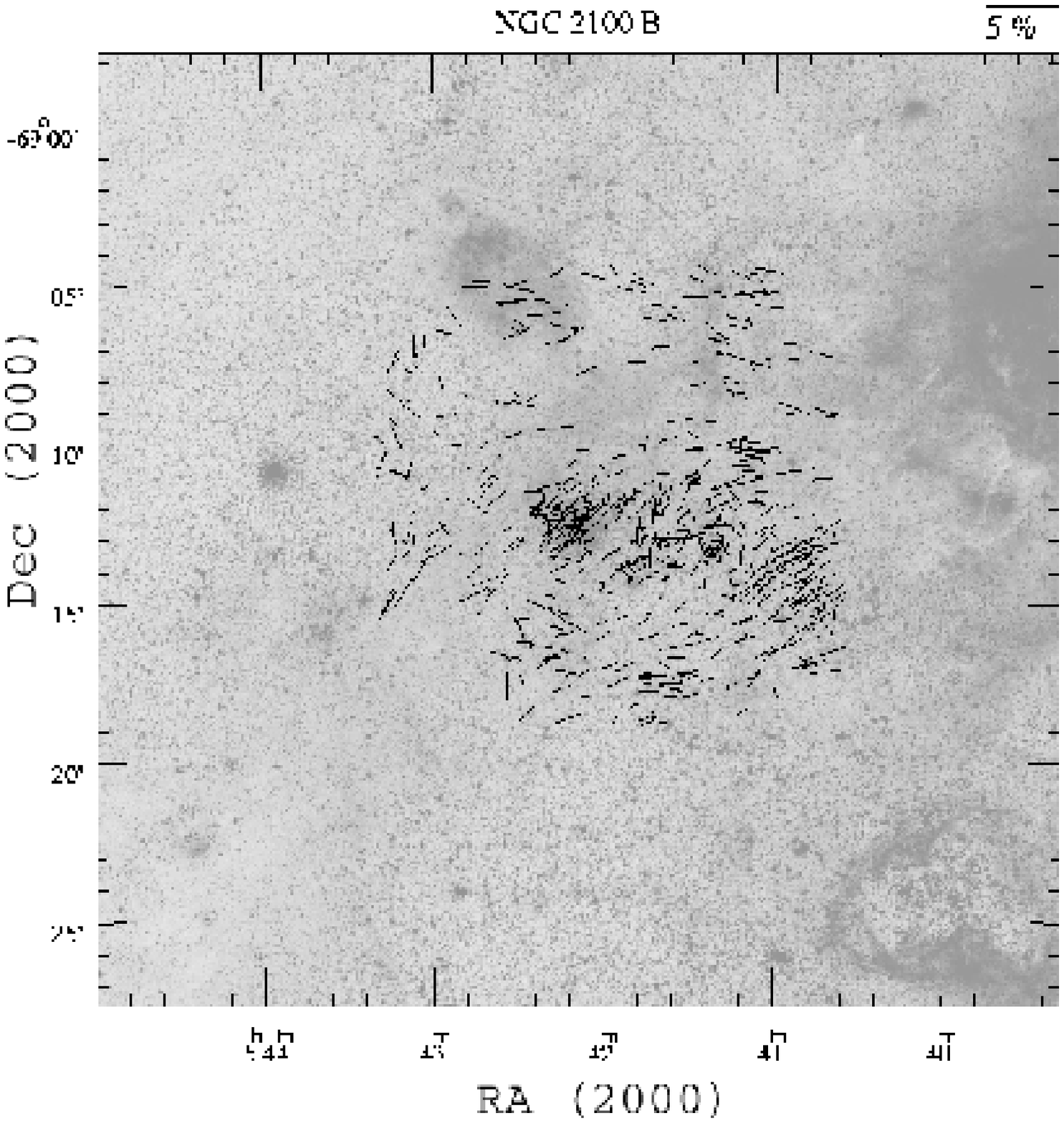}
\caption{B filter polarization of 592 targets which have 
p$/\sigma_{p} > 3$ in the vicinity of NGC 2100.  Polarization vectors
are overplotted on a 0.5 deg$^{2}$ DSS2 blue image.}
\end{figure}

\begin{figure}
\includegraphics[scale=0.5]{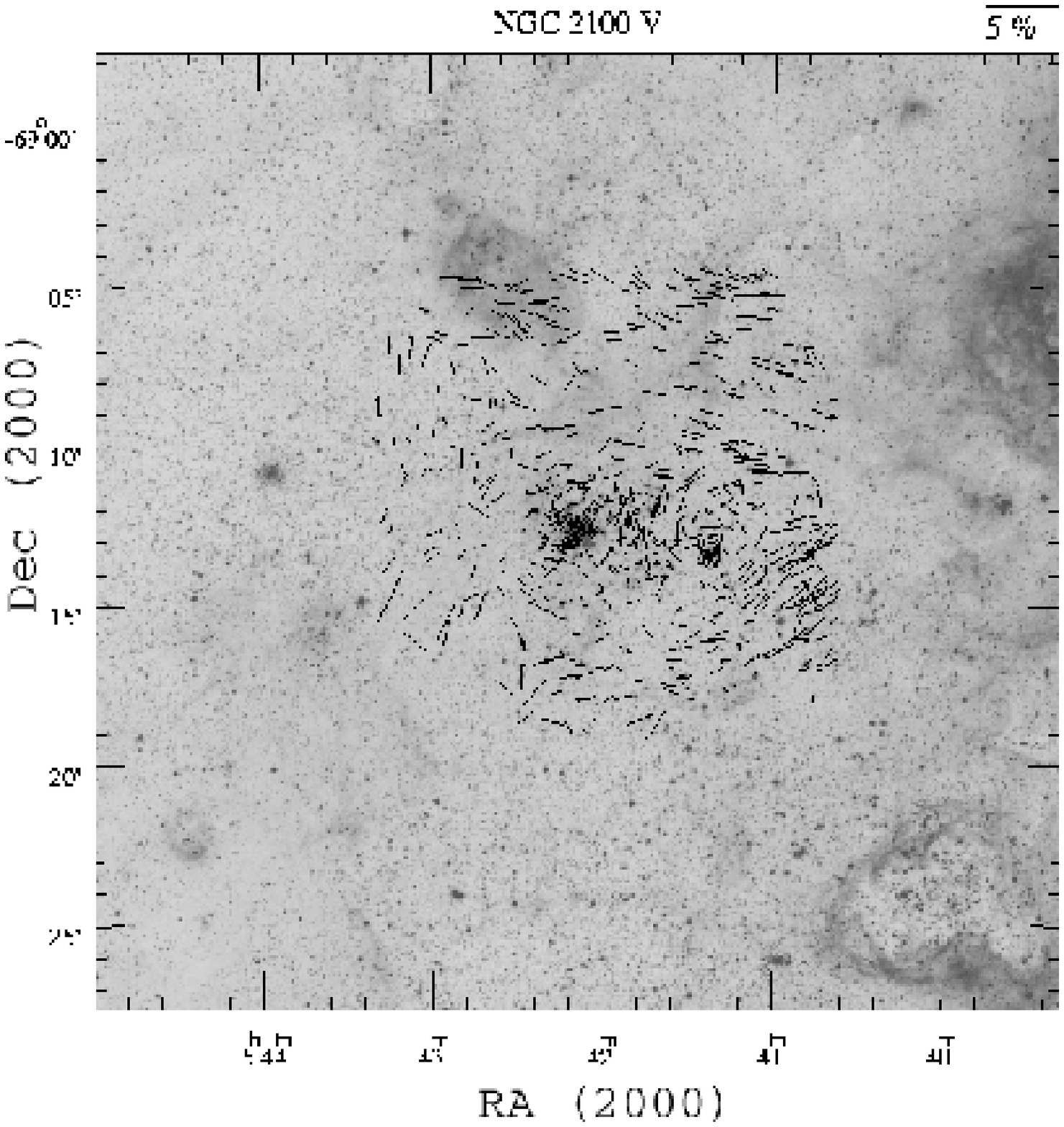}
\caption{V filter polarization of 661 targets which have 
p$/\sigma_{p} > 3$ in the vicinity of NGC 2100.  Polarization vectors
are overplotted on a 0.5 deg$^{2}$ DSS2 red image.}
\end{figure}

\begin{figure}
\includegraphics[scale=0.5]{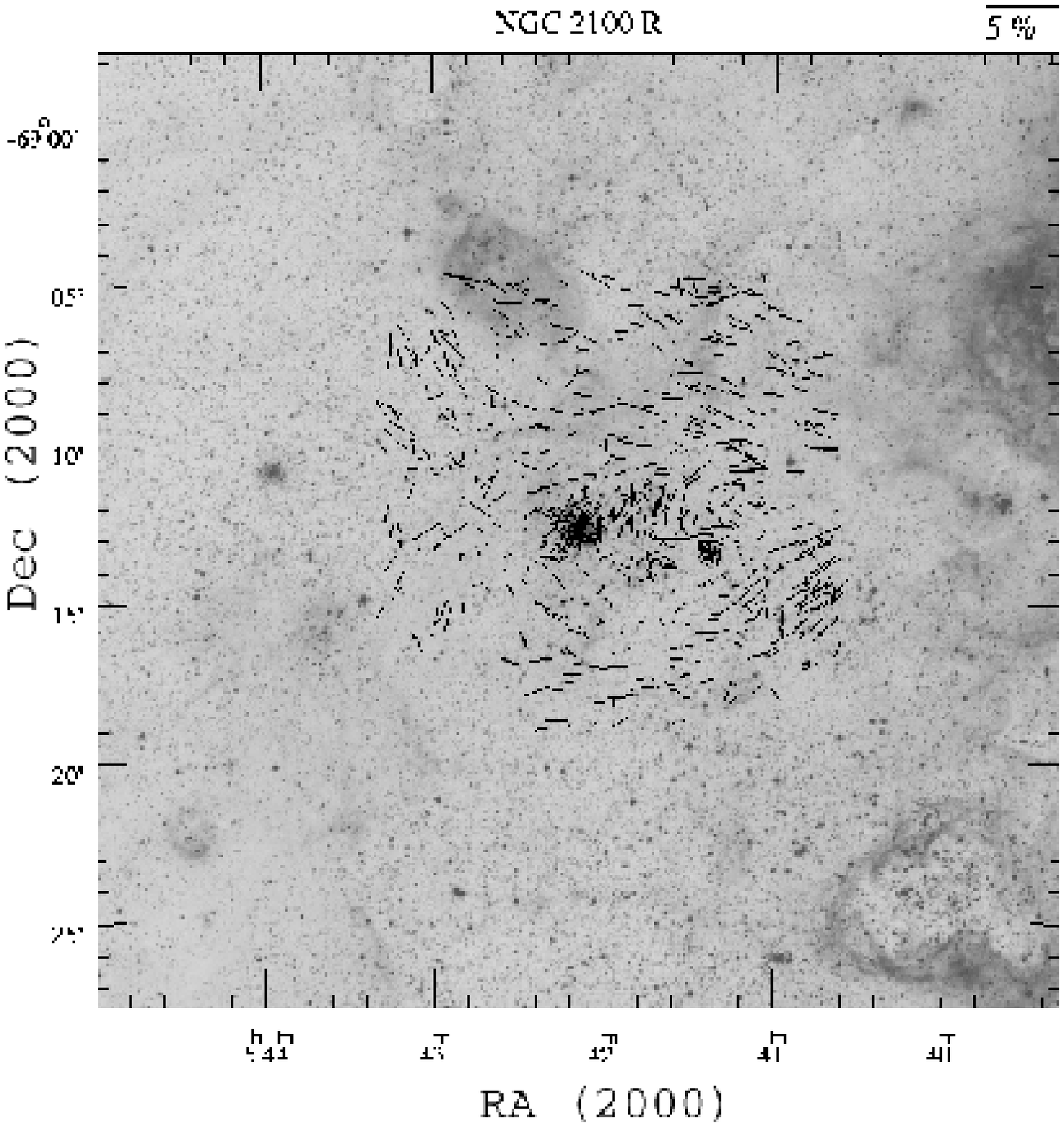}
\caption{R filter polarization of 700 targets which have 
p$/\sigma_{p} > 3$ in the vicinity of NGC 2100.  Polarization vectors
are overplotted on a 0.5 deg$^{2}$ DSS2 red image.}
\end{figure}

\begin{figure}
\includegraphics[scale=0.5]{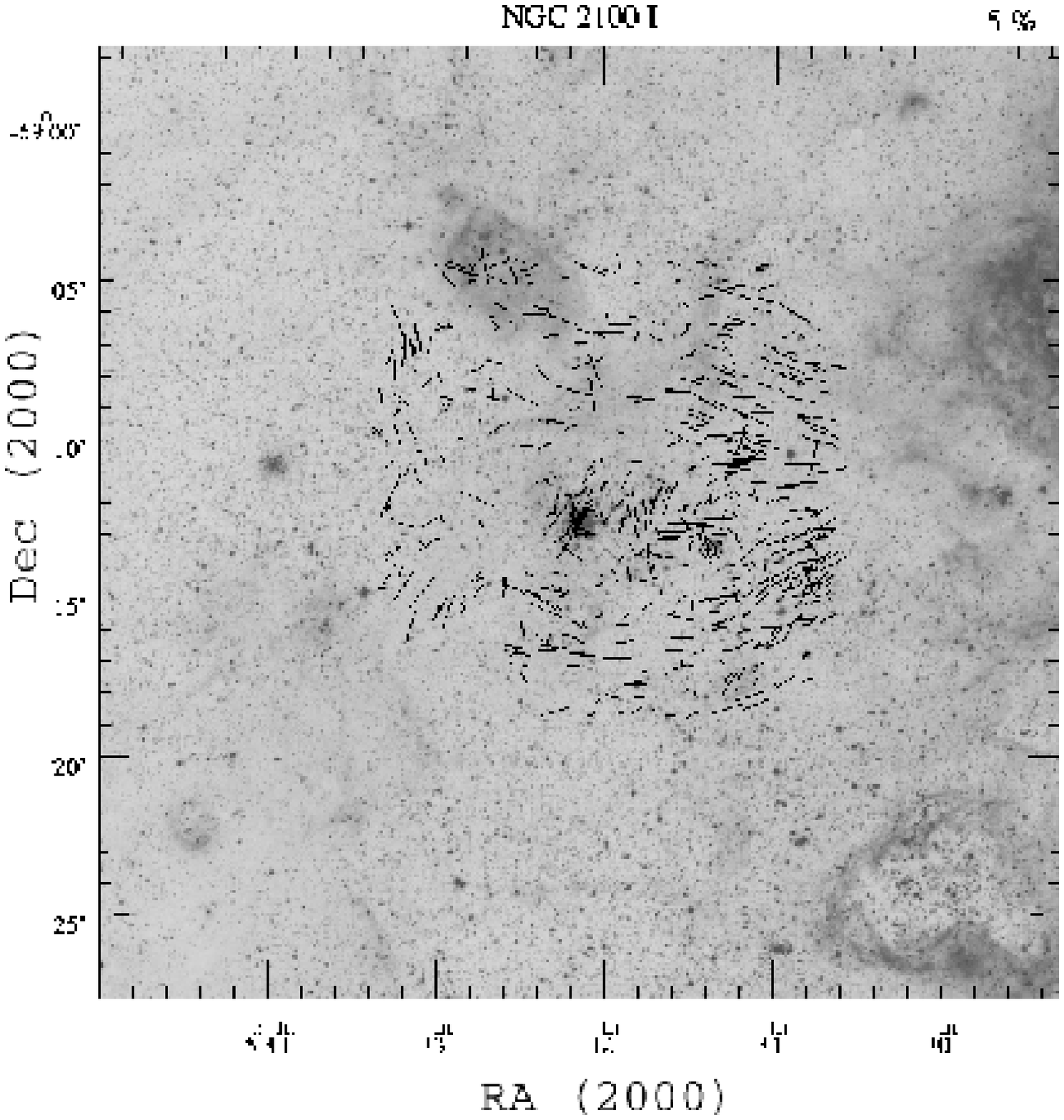}
\caption{I filter polarization of 621 targets which have 
p$/\sigma_{p} > 3$ in the vicinity of NGC 2100.  Polarization vectors
are overplotted on a 0.5 deg$^{2}$ DSS2 red image.}
\end{figure}

\begin{figure}
\includegraphics[scale=0.4]{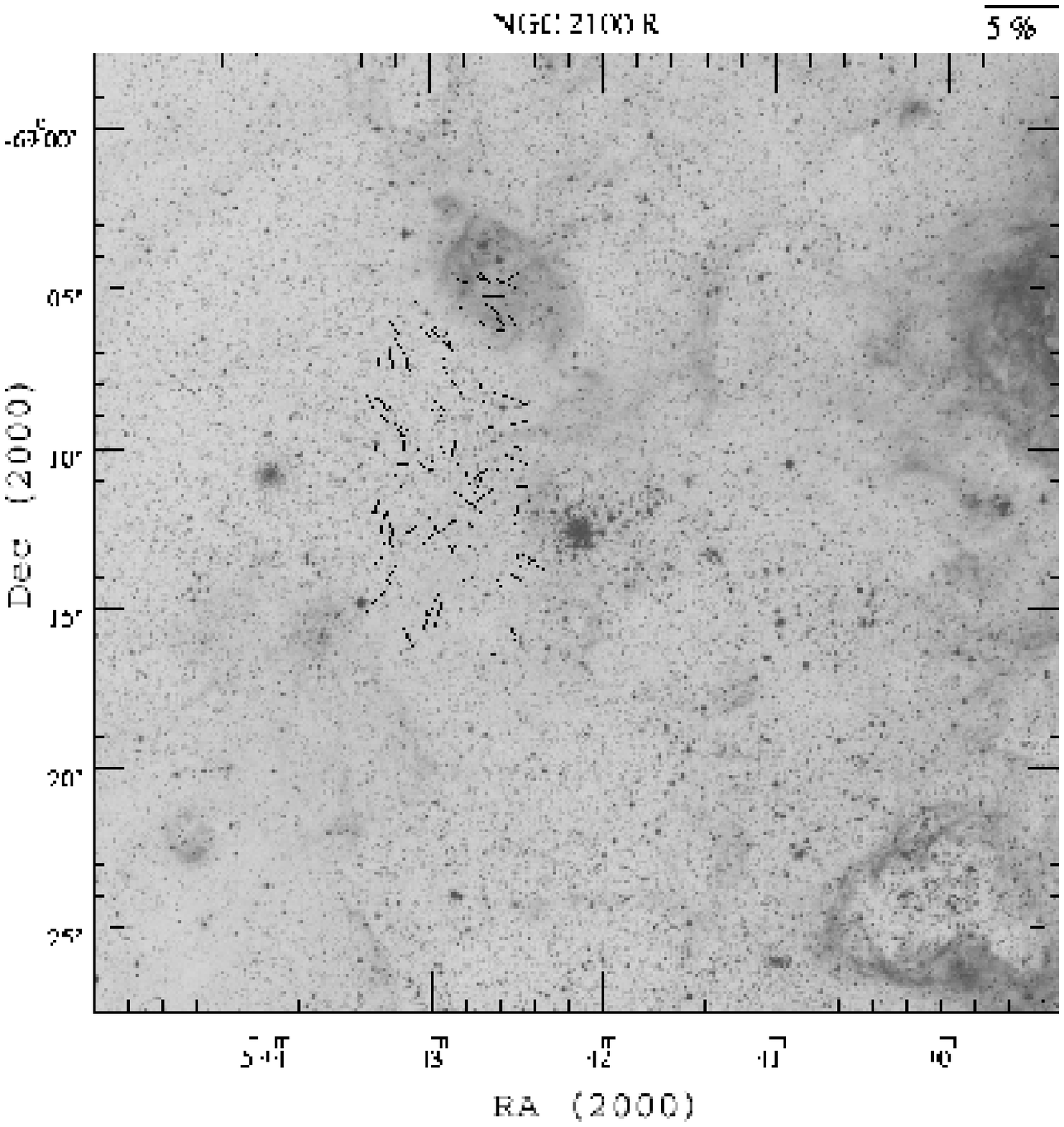}
\includegraphics[scale=1.0]{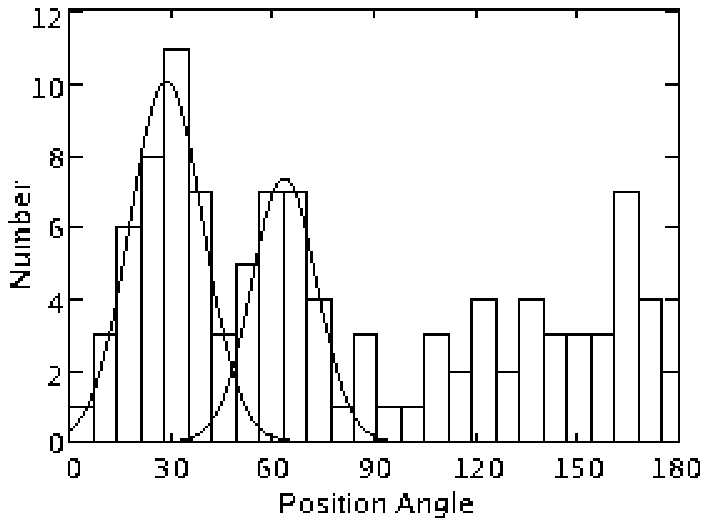}
\caption{Section A of our field of view, showing the polarization of 105
objects with 0.1\% $<$ p $<$ 3.0\% and p$/\sigma_{p} > 3$.  We find the
position angles in this region are concentrated at angles $<$ 90$^\circ$,
possibly following a bimodal distribution with centers at 29$^\circ$ and
64$^\circ$.}
\end{figure}

\begin{figure}
\includegraphics[scale=0.4]{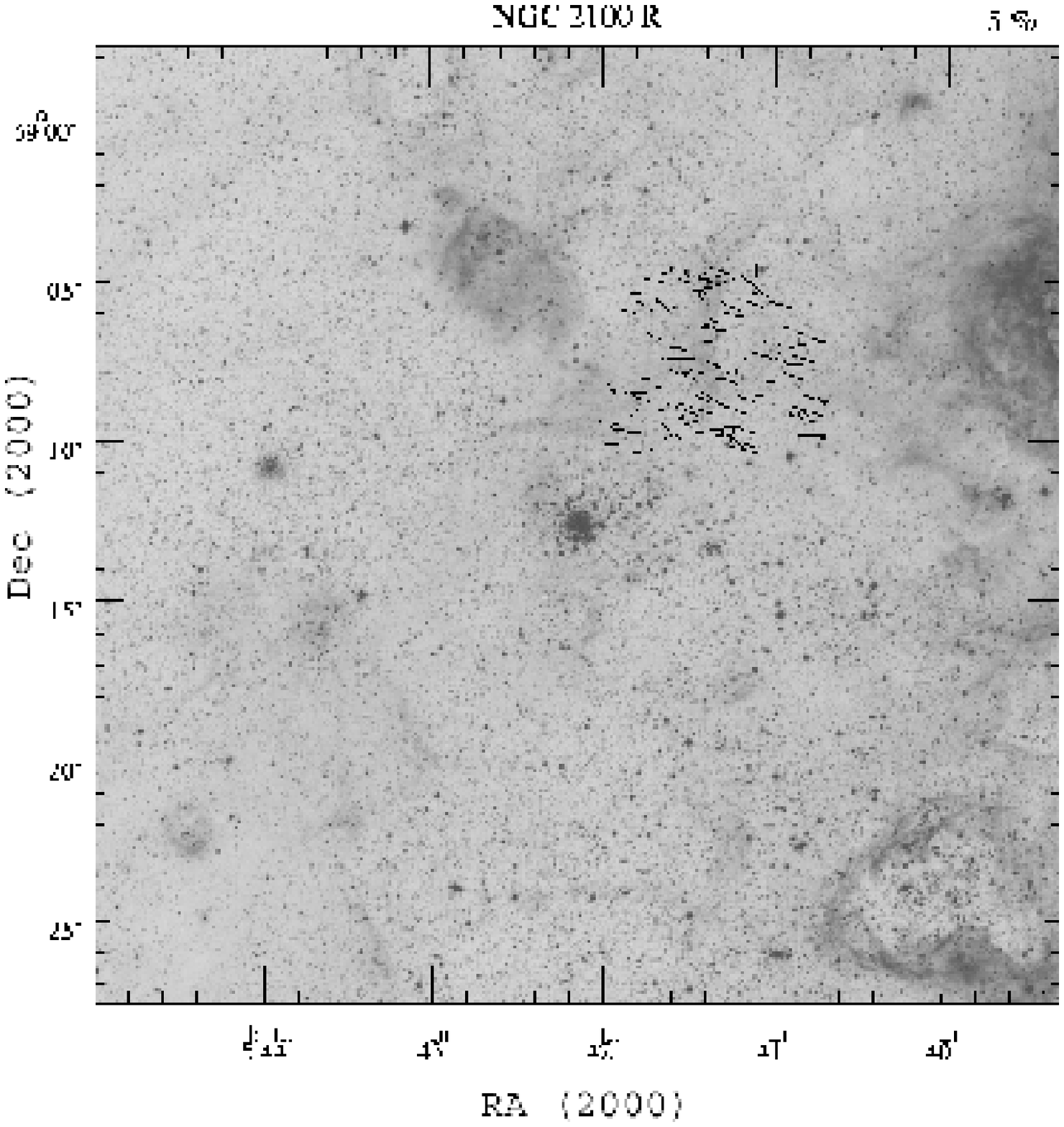}
\includegraphics[scale=0.29]{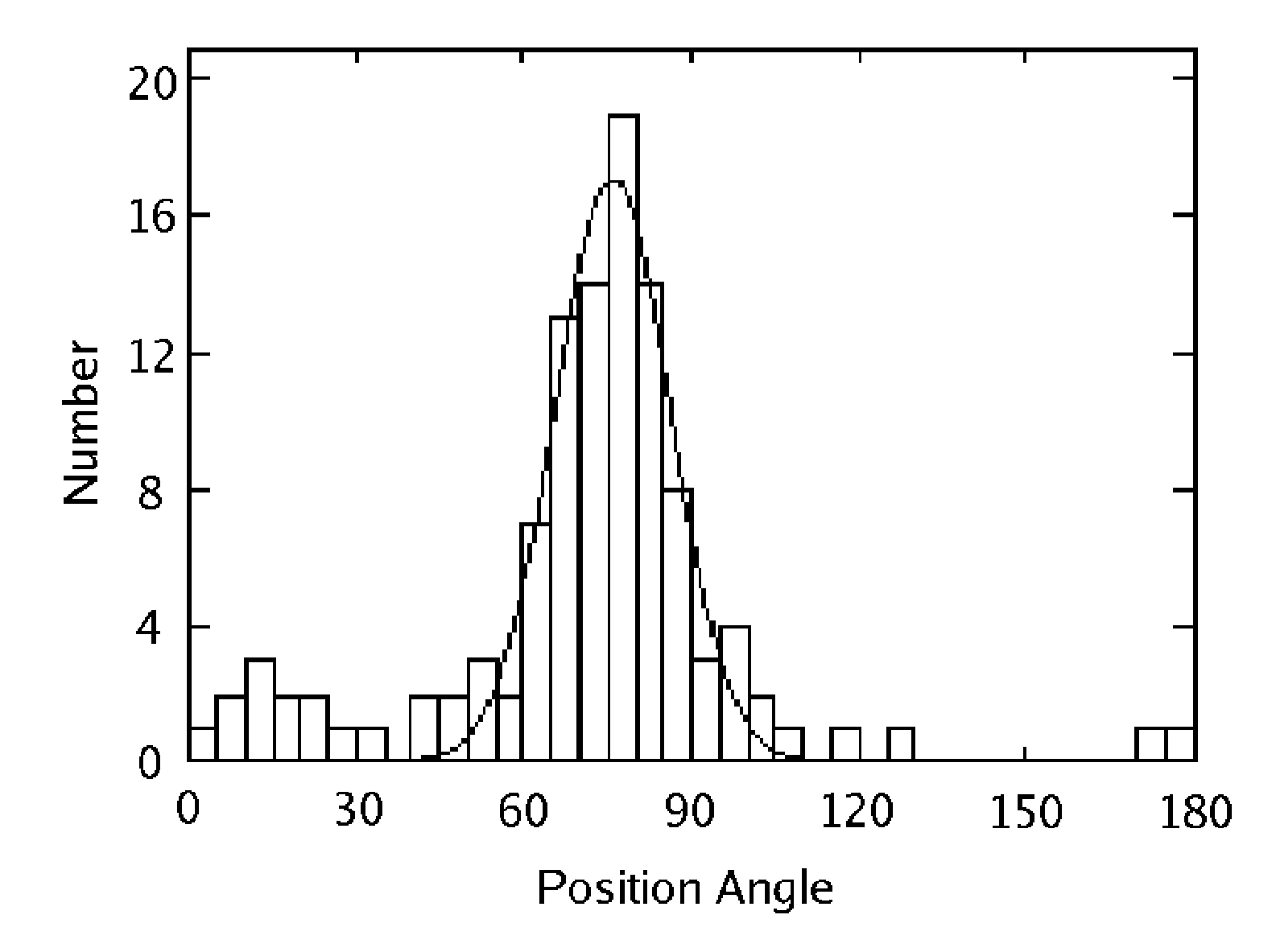}
\caption{Section B of our field of view, showing the polarization of 110
objects with 0.1\% $<$ p $<$ 3.0\% and p$/\sigma_{p} > 3$.  The mean 
polarization position angle of these stars, determined by a gaussian fit,
is 76$^{\circ}$.} 
\end{figure}

\begin{figure}
\includegraphics[scale=0.4]{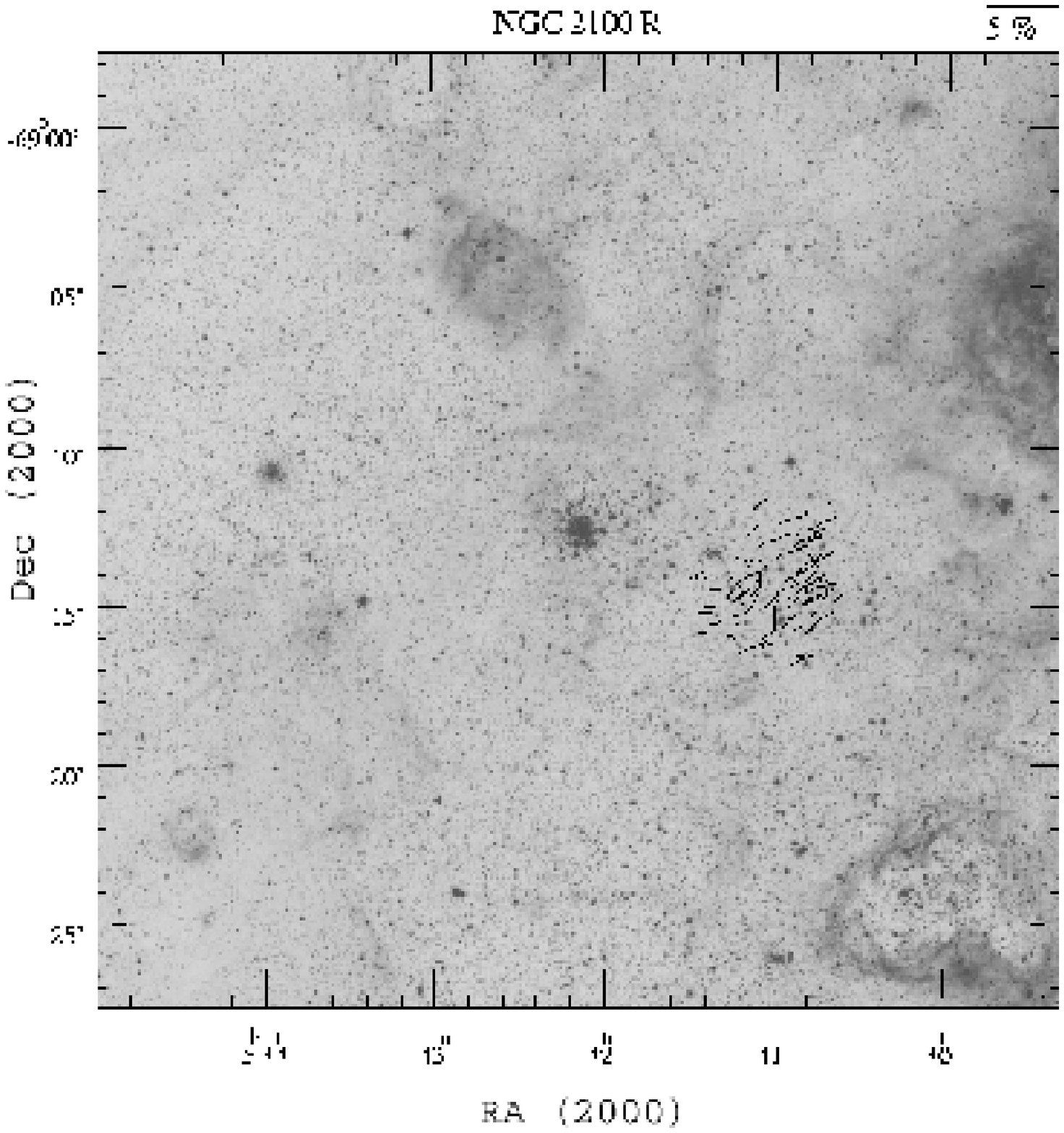}
\includegraphics[scale=0.7]{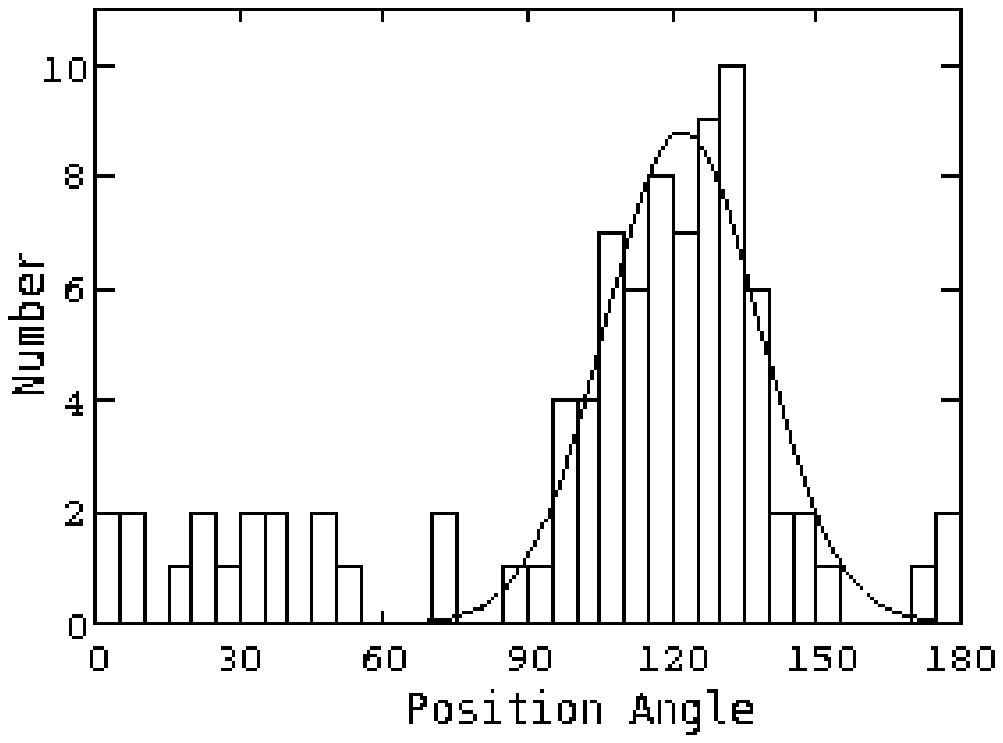}
\caption{Section C of our field of view, showing the polarization of 88
objects with 0.1\% $<$ p $<$ 3.0\% and p$/\sigma_{p} > 3$.  The mean 
polarization position angle of this region was determined to be 122$^{\circ}$.}
\end{figure}

\begin{figure}
\includegraphics[scale=0.4]{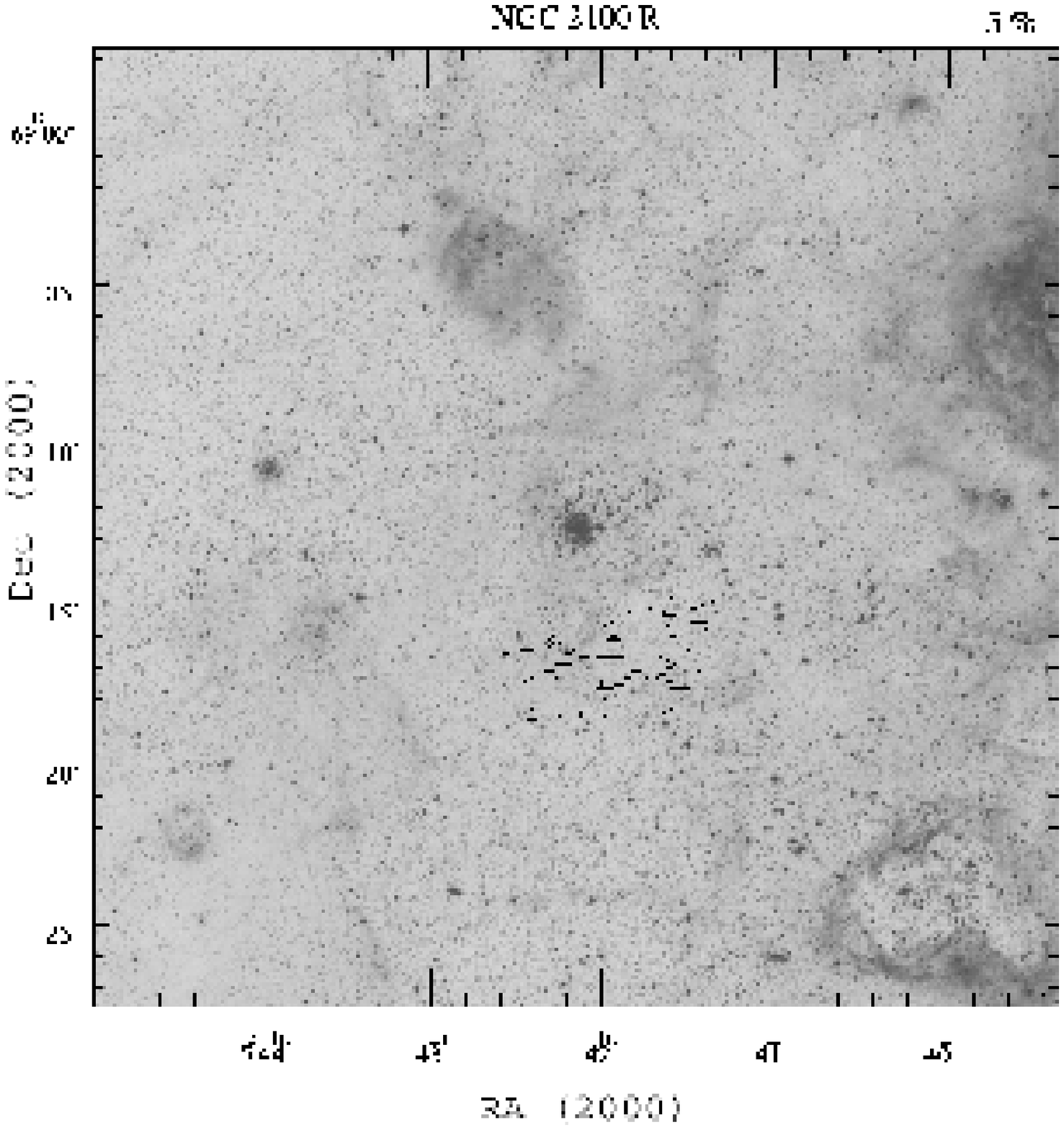}
\includegraphics[scale=1.0]{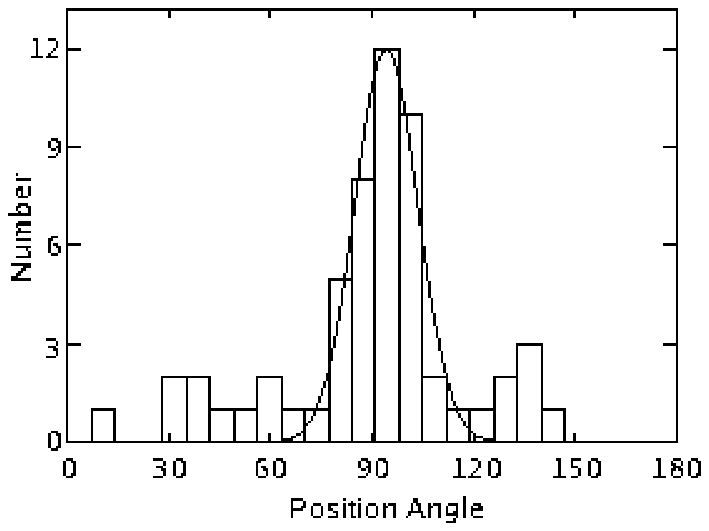}
\caption{Section D of our field of view, showing the polarization of 56
objects with 0.1\% $<$ p $<$ 3.0\% and p$/\sigma_{p} > 3$.  The mean 
polarization position angle of this region was determined to be 94$^{\circ}$.}
\end{figure}

\begin{figure}
\includegraphics[scale=0.4]{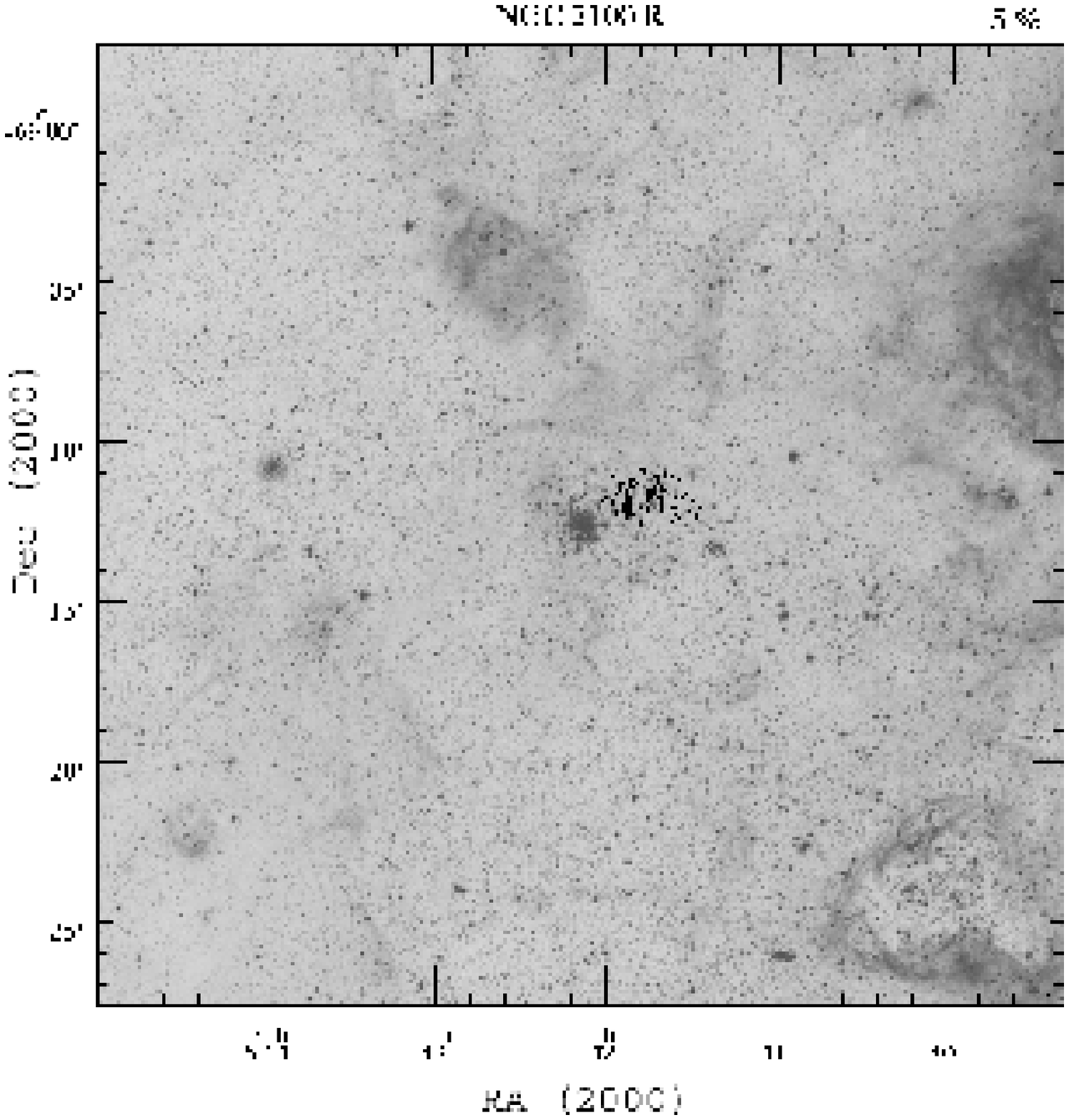}
\includegraphics[scale=1.0]{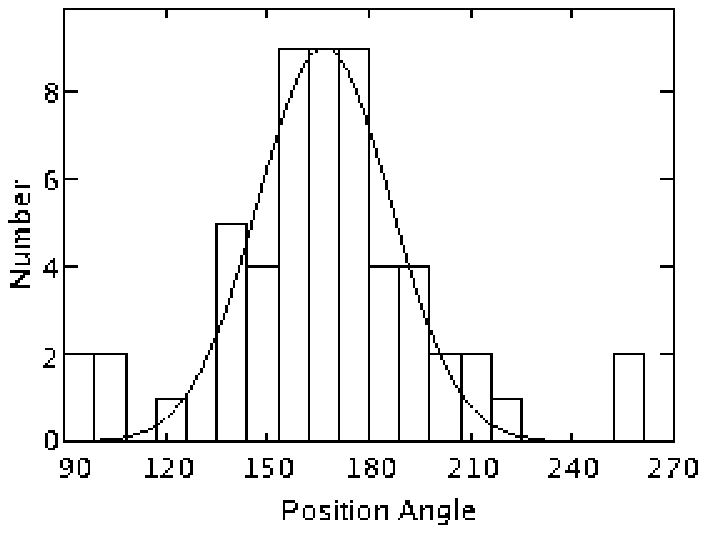}
\caption{Section E of our field of view, showing the polarization of 56
objects with 0.1\% $<$ p $<$ 3.0\% and p$/\sigma_{p} > 3$.  The mean
polarization position angle of this region was determined to be 167$^{\circ}$.}
\end{figure}

\end{document}